\documentclass[aps,prd,amsmath,amssymb,superscriptaddress,showkeys,showpacs,twocolumn,floatfix]{revtex4-1}

\usepackage[final]{graphicx}
\usepackage{hyperref}
\usepackage{amsmath}
\usepackage{bbm}
\usepackage{amsfonts}
\usepackage{amssymb}
\usepackage{latexsym}
\usepackage{graphicx}
\usepackage[english]{babel}
\usepackage{multirow}
\usepackage{float}
\usepackage{url}
\usepackage{slashed}
\usepackage{xcolor}
\usepackage[utf8]{inputenc}
\usepackage{verbatim}
\usepackage{lipsum}

\usepackage[utf8]{inputenc}

\usepackage{diagbox}

\usepackage{mathtools}
\usepackage{amsfonts}
\usepackage{mathrsfs}
\usepackage{bbm}
\usepackage{slashed}

\usepackage{graphicx}
\usepackage{color}
\usepackage{array}
\usepackage{esint}
\usepackage{placeins}
\usepackage{booktabs}
\usepackage{makecell}
\usepackage{epstopdf}
\usepackage[caption=false]{subfig}

\usepackage{xspace}
\usepackage{siunitx}
\usepackage{hyperref}
\usepackage[nameinlink]{cleveref}
\usepackage{appendix}

\usepackage{xifthen}
\usepackage{xcolor}

\hypersetup{colorlinks,linkcolor={black!75!black},citecolor={blue!75!black},urlcolor={blue!75!black}}

\usepackage{comment}

\usepackage{xcolor}

\usepackage{ulem}

\newcommand{\Tint}[1]{{\hbox{$\sum$}\!\!\!\!\!\!\!\int\,}_{\!\!\!\!\raise-0.9ex\hbox{$\scriptstyle{#1}$}}}

\newcommand{\lp}{\left(}
\newcommand{\rp}{\right)}

\begin{document}
\title{Theoretical Uncertainties in First-Order Electroweak Phase Transitions}

\author{Yongzhao Zhu}
\affiliation{Department of Physics and Chongqing Key Laboratory for Strongly Coupled Physics, Chongqing University, Chongqing 401331, P. R. China}

\author{Jie Liu}
\affiliation{Department of Physics and Chongqing Key Laboratory for Strongly Coupled Physics, Chongqing University, Chongqing 401331, P. R. China}

\author{Renhui Qin}
\email{20222701021@stu.cqu.edu.cn}
\affiliation{Department of Physics and Chongqing Key Laboratory for Strongly Coupled Physics, Chongqing University, Chongqing 401331, P. R. China}

\author{Ligong Bian}
\email{lgbycl@cqu.edu.cn}
\affiliation{Department of Physics and Chongqing Key Laboratory for Strongly Coupled Physics, Chongqing University, Chongqing 401331, P. R. China}

\date{\today}

\begin{abstract}
We systematically investigate theoretical uncertainties in perturbative analyses of first-order electroweak phase transitions (EWPT). Utilizing the Standard Model Effective Field Theory (SMEFT) framework, we quantify the gauge dependence, renormalization scheme, and scale dependences in predicting phase-transition parameters across both zero and finite chemical potential regimes. Our key findings reveal that: 1)
the gauge-parameter dependence and the chemical potential are subdominants, and 2) Baryon number preservation criteria and phase transition observables exhibit pronounced sensitivity to the renormalization scale in the ${\overline{\rm MS}}$ scheme.
Comparative analyses with the on-shell scheme demonstrate that within strongly first-order EWPT parameter spaces, the ${\overline{\rm MS}}$ scheme predicts enhanced phase transition strengths and prolonged transition durations.
\end{abstract}

\maketitle

\section{Introduction}

The persistent null results in collider searches for beyond Standard Model (BSM) particles over multiple decades strongly suggest two plausible scenarios: either these hypothetical states possess exceptionally large masses (exceeding current collider energy thresholds) or exhibit extremely feeble couplings to Standard Model (SM) sector. This observational landscape has driven two distinct methodological branches in particle phenomenology: (i) Top-down model building constrained by mass/coupling hierarchies, where new physics (NP) constructions (e.g., supersymmetry, composite Higgs) inherently satisfy these observational limits;
(ii) Bottom-up effective theory approaches, particularly through the Standard Model Effective Field Theory (SMEFT) framework \cite{Datta:2025csr}, which systematically parametrizes heavy NP effects via higher-dimensional operators when
$\Lambda_{NP}\gg v_{EW}$ (where
$v_{EW}\approx 246$ GeV denotes the electroweak scale). This framework respects the Standard Model (SM) gauge symmetry $SU(3)_c\bigotimes SU(2)_L\bigotimes U(1)_Y$ and incorporates only SM particles. A barrier between the symmetric and broken phases is introduced at the tree-level potential via a dimension-six operator $(\Phi^\dagger\Phi)^3/\Lambda^2$
 where $\Lambda$ serves as the sole new physics (NP) scale in the theory. This barrier facilitates a first-order phase transition (PT), offering a critical mechanism to probe the universe’s evolution. Such transitions, typically realized in SM extensions, can account for the baryon asymmetry through electroweak baryogenesis~\cite{Morrissey:2012db} and generate observable gravitational wave (GW) signals~\cite{Caldwell:2022qsj,Athron:2023xlk,Bian:2021ini}.

Previous studies of the EWPT within the SMEFT framework employ four-dimensional finite-temperature thermal field theory. These analyses demonstrate that extensive regions of parameter space can generate a strong first-order PT, detectable by near-future GW observatories~\cite{Ellis:2018mja,Cai:2017tmh,Croon:2020cgk,Grojean:2004xa,Delaunay:2007wb,Chala:2018ari,Bodeker:2004ws,Damgaard:2015con,Hashino:2022ghd,Postma:2020toi}. Most of these investigations adopt either the on-shell (${\rm OS}$) or ${\overline{\rm MS}}$ renormalization scheme in the Landau gauge~\cite{Cai:2017tmh,Croon:2020cgk,Grojean:2004xa,Delaunay:2007wb,Bodeker:2004ws,Hashino:2022ghd,Damgaard:2015con,Chala:2018ari}.
The condition $v_C/T_C\gtrsim 1$, where the $v_C$ denotes the Higgs vacuum expectation value at the critical temperature $T_C$,
has been conventionally adopted as the criterion for a strongly first-order phase transition (SFOPT). Recent work by Ref.~\cite{Athron:2022jyi} highlighted significant theoretical uncertainties arising from gauge dependence and renormalization scale variations in calculations of phase transition parameters within the Standard Model (SM) extended by a real scalar singlet. Motivated by these findings, we systematically investigate the sensitivity of the SFOPT criteria to renormalization scheme and scale choices, quantitatively evaluate gauge artifacts in the $R_\xi$ gauge framework, and establish robust SFOPT criteria based on baryon number washout avoidance conditions.

This paper is organized as follows. In Section~\ref{secpotential}, we review the finite-temperature effective potential formalism and address challenges in maintaining manifest gauge invariance, particularly through the Nielsen identity.
Section~\ref{secPT} systematically examines the PT dynamics, including the effect of gauge choice and finite-density corrections (chemical potential effects), scheme-dependent discrepancies between the ${\rm OS}$ and ${\overline{\rm MS}}$ scheme, and the renormalization scale dependence inherent to the. Finally, in Section~\ref{seccon}, we summarize our findings and discuss their implications for future studies on PT.

\section{Thermal effective potential}\label{secpotential}

We consider the standard mode effective field theory(SMEFT) with massless leptons. The Lagrangian is
\begin{equation}\label{lagrangian}
\begin{aligned}
\mathcal{L}=&-\frac{1}{4}F_{\mu\nu}F_{\mu\nu}-\frac{1}{4}B_{\mu\nu}B_{\mu\nu}+(D_\mu \Phi)^\dagger(D_\mu \Phi)\\
&+\mu_h^2\Phi^\dagger \Phi-\lambda(\Phi^\dagger \Phi)^2-c_6(\Phi^\dagger \Phi)^3\\
&+\overline{e}_R(i\slashed{\partial}-g^\prime \slashed{B})e_R+y_t(\overline{q}_L\tilde{\Phi}t_R+\overline{t}_R\Tilde{\Phi}^\dag q_L)\\
&+\overline{\psi}_L(i\slashed{\partial}-\frac{1}{2}g^\prime\slashed{B}+\frac{1}{2}g \tau ^i A^i)\psi_L,
\end{aligned}
\end{equation}
where, $\psi_L=(\nu_L,e_L)^T$ and $ c_6=1/\Lambda^2$ with $\Lambda$ being the the NP energy scale. The covariant derivative $D_\mu$ is defined as $D_\mu=\partial_\mu-i g T^i A^i_\mu-ig^\prime B_\mu$. The $F_{\mu\nu}=\partial_\mu A_\nu-\partial_\nu A_\mu$ and $B_{\mu\nu}=\partial_\mu B_\nu-\partial_\nu B_\mu$ are the  field strength tensors of $SU(2)_L$ and $U(1)_Y$. We consider the top quark has non-zero mass with Yukawa coupling $y_t$, and $\tilde{\Phi}=i\sigma_2 \Phi^*$, where $\Phi$ is the SM Higgs doublet.

The electromagnetic and weak-neutral charges of this gauge theory as defined as
\begin{align}
N_1&=\int d^3 x (j_0+j_0^3),\\
N_3&=\int d^3x (\sin^2\theta j_0-\cos^2\theta j_0^3)\frac{2}{\cos 2\theta}
\end{align}
and the lepton number is conserved
\begin{equation}
N_2=\int d^3x (\overline{e}\gamma_0 e+\frac{1}{2}\overline{\nu}\gamma_0(1+\gamma_5)\nu)
\end{equation}
where $j_0,j_0^3$ are the associated currents \cite{Kapusta:1990qc}. The partition function is obtained by introducing the chemical potential $\mu_i$
\begin{equation}
\begin{aligned}
Z=&\text{Tr} e^{-\beta(H-\sum_i\mu_i N_i)}\\
=&\int d\varphi exp\left[\int_0^\beta d t \int d^3x (\tilde{\mathcal{L}}+\mu_2\overline{e}\gamma_0 e\right.\\
&\left.+\mu_2\frac{1}{2}\overline{\nu}\gamma_0(1+\gamma_5)\nu)\right].
\end{aligned}
\end{equation}
where $\tilde{\mathcal{L}}$ is obtained from $\mathcal{L}$ by \cite{Kapusta:1990qc}
\begin{equation}\label{newgaugefield}
\begin{aligned}
B_\mu &\rightarrow B_\mu-\left(\mu_1+\frac{2\sin^2\theta}{\cos 2\theta}\mu_3\right)\frac{1}{g^\prime}\delta_{\mu 0},\\
A_\mu &\rightarrow A_\mu-\left(\mu_1-\frac{2\cos^2\theta}{\cos2\theta}\mu_3\right)\frac{1}{g}\delta_{\mu 0}\delta^{i 3}.
\end{aligned}
\end{equation}

After considering the ghost term and the gauge parameter $\xi$, the final Lagrangian has the form
\begin{equation}
\begin{aligned}
\mathcal{L}_{eff}
=&\tilde{\mathcal{L}}+\mu_2\left[\overline{e}\gamma_0 e+\frac{1}{2}\overline{\nu}\gamma_0(1+\gamma_5)\nu\right]+\tilde{\mathcal{L}}_{ghost}\\
&-\frac{1}{2\xi}(\partial^\mu A_\mu^a-\frac{1}{2}\xi g \phi \chi_i)^2\\
&-\frac{1}{2\xi}(\partial^\mu B_\mu-\frac{1}{2}\xi g^\prime \phi \chi_3)^2
\end{aligned}
\end{equation}
where $\chi_i$ is goldstone field and $\phi$ is background field. The $\tilde{\mathcal{L}}_{ghost}$ is the common contribution of ghost field $\eta$ but the gauge fields are replaced by Eq.\eqref{newgaugefield}.
%$\omega=-T\partial \ln Z/\partial V$
The tree-level thermal potential  is obtained by setting all the fields of $\mathcal{L}_{eff}$ to zero except $\phi$
\begin{equation}\label{vefftree}
V_{tree}=-\frac{1}{2}(\mu_h^2+\left(\frac
{\mu_3}{\cos 2\theta}\right)^2)\phi^2+\frac{1}{4}\lambda \phi^4+\frac{1}{8}c_6\phi^6.
\end{equation}
 We adopt the 4d Daisy resummation to study the electroweak PT at one-loop order, where the one-loop thermal potential has the form of (see. Appendix.\ref{appendixoneloop})
\begin{equation}
V_{1loop}=V_{CW}+V_{T}+V_{Daisy}\;.
\end{equation}
Where, the one-loop zero-temperature Coleman-Weinberg potential $V_{CW}$ in the ${\rm OS}$ scheme and ${\overline{\rm MS}}$ scheme are
\begin{equation}\label{VCW}
\begin{aligned}
V_{CW}^{\rm OS}=&\sum_{i} n_i\frac{1}{64\pi^2}\bigg[m_i^4\left(\log\left(\frac{m_i^2}{m^2_{i,v}}\right)-\frac{3}{2}\right)\\
&+2m_i^2 m^2_{i,v}\bigg]\;,\\
V_{CW}^{\overline{MS}}=&\sum_{i}n_i\frac{m_i^4}{64\pi^2}\left(\log\left(\frac{m_i^2}{\overline{\mu}^2}\right)-c\right)\;,%\quad z=\frac{m}{T}
\end{aligned}
\end{equation}
where $i=\{\phi,\chi_{1},\chi_2,\chi_3,\,W,Z,t,\eta_c,\eta_0\}$ and $c=5/6$ for bosons and $3/2$ for fermion, the $m_{i,v}$ is mass which calculated at VEV ($v=246.22$GeV). The $n_i$ is the degrees of freedom, $n_{\{\phi,\chi_1,\chi_2,\chi_3,\,W\,Z\,t,\eta_c,\eta_0\}}=\{1,1,1,1,6,3,-12,2,1\}$. The chemical potential affects the scalar and $W^{\pm}$ boson mass, and the gauge parameter enters the ghost and goldstone mass.
We set the $\overline{\mu}=246.22$ GeV, $\overline{\mu}=T$ to study the renormalization scale dependence in the ${\overline{\rm MS}}$ scheme.
The thermal piece of the one-loop thermal potential is~\cite{Bernon:2017jgv}:
\begin{equation}\label{VT}
\begin{aligned}
V_T(z)=&\sum_{i}n_i \frac{T^4}{2\pi^2} J_{b,f}\\
=&\mp\sum_{i}n_i \frac{T^4}{2\pi^2}\left(\sum_{l=1}^n \frac{(\pm 1)^l}{l^2}\left(\frac{m_i
}{T}\right)^2 K_2\left(\frac{m_i
}{T} l\right)\right)\;,
\end{aligned}
\end{equation}
with $K_2$ being the second order of modified Bessel function, and we take $n=7$ in this work.
The final thermal effective potential at one-loop level is
\begin{equation}\label{veff}
V_{eff}=V_{tree}+V_{1loop}\;.
\end{equation}

\begin{figure}[!htp]
    \centering
\includegraphics[width=0.8\linewidth]{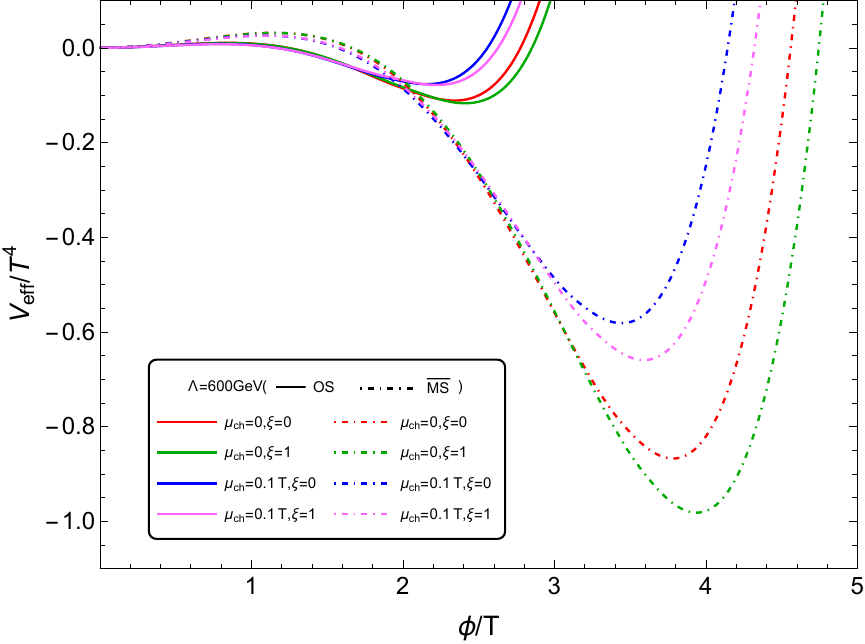}
    \caption{The effective potential $V_{CW}^{\overline{MS}} (\bar{\mu}=246.22\text{GeV})$ and $V_{CW}^{OS}$ with $\Lambda=600$GeV at temperature $T_n$. }
    \label{figveff}
\end{figure}

Since the perturbation of 4d effective potential will break down in the Goldstone field with a large $\xi$, we consider the gauge parameter satisfies the condition $|\xi| \ll 1/g$~\cite{Croon:2020cgk}. On the other hand, the lepton chemical potential will change the W boson mass, it needs less than the $m_W$ since the number of bosons must be positive~\cite{Haber:1981fg}.
Taking the relation $\mu_3=\mu_{ch},\mu_1=16/11\mu_{ch}$~\cite{Kapusta:1990qc}, we show that the potential energy in the ${\rm OS}$ scheme is larger than that in the ${\overline{\rm MS}}$ scheme in Fig.\ref{figveff}, and we found that this difference increased with $\Lambda$ increasing. As will be explored later, the effects of the gauge parameter and the chemical potential on the thermal effective potential drive the uncertainties in the PT parameters predictions.

The effective potential is gauge-dependent since the goldstone and ghost terms appear at the one-loop level. This dependence led to a theoretical uncertainty in the predictions of PT parameters.
A gauge-independent potential can be constructed considering the Nielsen identity~\cite{Nielsen:1975fs,Kobes:1990dc,Metaxas:1995ab,Garny:2012cg,DiLuzio:2014bua}
\begin{align}\label{Nielsen}
\xi\frac{\partial V_{\rm eff}(\phi,\xi,T)}{\partial \xi}+C\frac{\partial V(\phi,\xi,T)}{\partial \phi}
=0\;,
\end{align}
 which means the $\xi$ dependence in the effective potential can be compensated by the $\xi$ dependence in the background field, i.e.,  $C=\xi\partial\phi/\partial\xi$. This equation works when the effective potential is gauge-independent at the true vacuum. One can construct a theory of gauge invariance with the exclusion of the Daisy term. The effective potential at the one-loop level becomes
\begin{equation}\label{GIveff}
V_{eff}=V_{tree}+V_{CW}+V_T.
\end{equation}
With the power counting $\lambda\sim g^2,c_6\sim g^4/T^2$, the effective potential is split to different order
\begin{equation}
\begin{aligned}
   V_{eff}&=V_{g^2}+V_{g^3}+V_{g^4}+...\\
   C&=C_{g}+C_{g^2}+C_{g^3}+...
\end{aligned}
\end{equation}
The Nielsen identity Eq.\ref{Nielsen} at $\mathcal{O}(g^4)$ has the form
\begin{equation}
\xi\frac{\partial V_{g^4}}{\partial\xi}=-C_{g^2}\frac{\partial V_{g^2}}{\partial \phi}.
\end{equation}
Here we include the $c_6(\Phi^\dagger\Phi)^3$ part in $V_{g^2}$ to create a barrier between the symmetric and broken phase at $\mathcal{O}(g^2)$, $V_{g^2}=V_{tree}$. The factor $C$ can be obtained by one loop thermally correction~\cite{Espinosa:2016nld}
\begin{equation}
\begin{aligned}
C_W(\phi,T,\xi)=&\frac{1}{2}g\Tint{k}\frac{\xi m_W}{(k^2-m_{\chi_{1}}^2)(k^2-m_{\eta_c}^2)},\\
C_Z(\phi,T,\xi)=&\frac{1}{2}g^\prime\Tint{k}\frac{\frac{1}{2}\xi m_B}{(k^2-m_{\chi_{3}}^2)(k^2-m_{\eta_0}^2)}\\
&+\frac{1}{2}g\Tint{k}\frac{\frac{1}{2}\xi m_W}{(k^2-m_{\chi_{3}}^2)(k^2-m_{\eta_0}^2)}\;,
\end{aligned}
\end{equation}
where $m_B=g^\prime\phi/2$. By defining the function
\begin{equation}
\begin{aligned}
   I_C(m_1,m_2)=\frac{1}{16\pi^2}&\left(\frac{T}{4\pi(m_1+m_2)}+L_b\right),\\
  L_b=\ln\left(\frac{\overline{\mu}^2}{T^2}\right)&+2\gamma_E-2\ln(4\pi)\;,
\end{aligned}
\end{equation}
with $\gamma_E$ is the Euler-Mascheroni constant, then
\begin{equation}
\begin{aligned}
C_{g^2}=&C_W(\phi,T,\xi)+C_Z(\phi,T,\xi)\\
=&\frac{1}{2}\bigg[\frac{1}{2}g(2m_W)I_C(m_{\chi_1},m_{\eta_c})\\
&+\frac{1}{2}g^\prime(\xi m_B)I_C(m_{\chi_3},m_{\eta_0})\\
&+\frac{1}{2}g(\xi m_W)I_C(m_{\chi_3},m_{\eta_0})\bigg],
\end{aligned}
\end{equation}
where the masses are listed in Eq.~\eqref{mass}.

  \begin{figure}[!htp]
    \centering
    \includegraphics[width=0.8\linewidth]{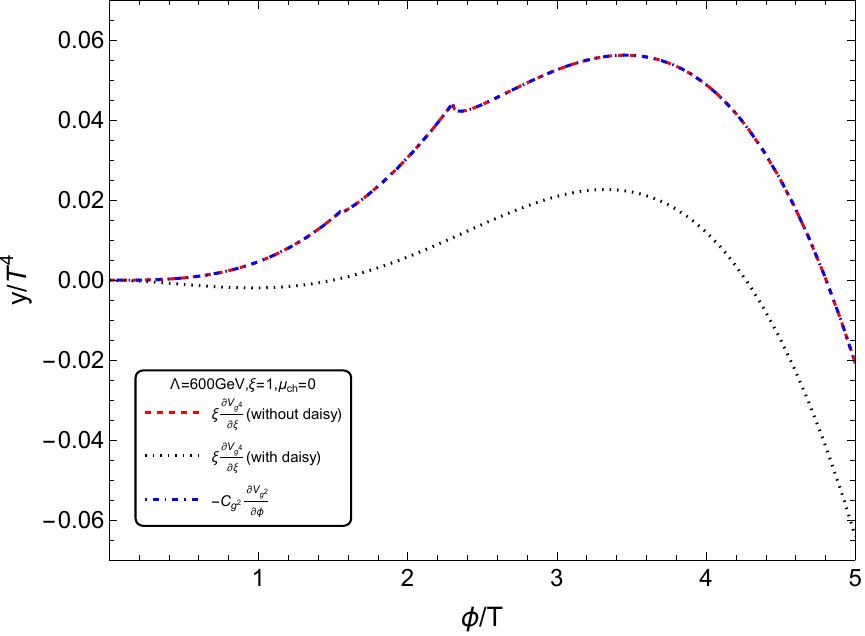}
    \caption{An example of the Nielsen identity with and without Daisy term at the temperature $T_n$ with $\Lambda=600$ GeV. The symbol $y$ denote the value of $\xi\partial V_{g^4}/\partial\xi$ and $-C_{g^2}\partial V_{g^2}/\partial \phi$. This figure is plotted by ${\overline{\rm MS}}$ scheme with $\overline{\mu}=246.22$ GeV, and the $V_T$ is calculated with high-temperature expansion for illustration.}
    \label{figGI}
\end{figure}

 The appearance of the Daisy term would modify the cubic term of $V_{T}$, and the Daisy part does not include the ghost mass, see the Appendix.~\ref{appendixoneloop} for details. Therefore, the $\xi$ dependent Daisy term in the thermal effective potential is not compatible with the Nielsen identity in the theoretical analysis, see also Ref.~\cite{Garny:2012cg} for the discussion in the abelian Higgs model. The Fig.~\ref{figGI} shows an example of Nielsen identity. The Nielsen identity is valid when the Daisy term is excluded from the effective potential, but it breaks when the Daisy part appears.  In the following sections, we investigate the gauge dependence in different renormalization schemes in detail.

\section{Phase transition dynamics}\label{secPT}

In this section, we study the effect of the gauge dependence and the chemical potential on the PT parameters in the popular renormalization scheme, i.e., ${\overline{\rm MS}}$ and ${\rm OS}$ schemes, and evaluate the scale dependence of PT parameters in the ${\overline{\rm MS}}$ scheme. We compute the GWs predicted by the SMEFT, and the range of NP scale $\Lambda$ that satisfies the constraints imposed by the strong first-order PT.

\subsection{PT parameters}

The Universe is in the ``symmetric'' phase at high temperature. As the temperature decreases to the critical temperature $T_c$, the ``symmetric'' and ``broken'' phases have the same free energy with
\begin{align}
&V_{eff}(\phi=\phi_C,T_C)=V_{eff}(\phi=0,T_C)\;,\nonumber\\
&\frac{dV_{eff}(\phi,T_C)}{d\phi}\big|_{\phi=\phi_C}=0\;.
\end{align}
 As the temperature continues to decrease, the energy of the ``broken'' phase is lower than that of the ``symmetric'' phase, which becomes metastable and decays to the broken phase through bubble nucleation. The nucleation temperature $T_n$ is obtained when the bubble nucleation rate is equal to Hubble parameter $\Gamma\sim H$ with $S_3/T\approx 140$~\cite{Linde:1981zj}.
 The Euclidean action is
\begin{equation}\label{action}
S_3=\int \bigg[\frac{1}{2}Z(\partial_i \phi)^2+V_{eff}(\phi,T) \bigg]d^3 x.
\end{equation}
where $Z$ is the renormalization factor of the wave function. The action can be obtained after the ``bounce solution" acquired
from the equation of motion:
\begin{equation}
Z\frac{d^2\phi}{d\rho^2}+Z\frac{2}{\rho}\frac{d\phi}{d\rho}+\frac{1}{2}Z^{\prime}\left(\frac{d\phi}{d\rho}\right)^2=\frac{dV_{eff}(\phi,T)}{d\phi}\;,
\end{equation}
with the boundary condition
\begin{equation}
\phi(\rho\rightarrow \infty)=0,\quad \left.\frac{d\phi}{d\rho}\right|_{\rho=0}=0\;.
\end{equation}
Since the factor $Z^{\prime}$ is close to zero at the minimal of the effective potential, we ignored the term $\frac{1}{2}Z^\prime(\frac{d\phi}{d\rho})^2$ in the bounce function, and the function become
\begin{equation}\label{bouncefun}
\frac{d^2\phi}{d\rho^2}+\frac{2}{\rho}\frac{d\phi}{d\rho}=\frac{1}{Z}\frac{dV_{eff}(\phi,T)}{d\phi}\;,
\end{equation}
and we use code ``findbounce'' to solve this equation and obtain the nucleation temperature $T_n$ and the corresponding background field value $\phi_n$~\cite{Guada:2020xnz}.
The inverse duration of the PT is defined as:
$\beta/H_n=T_n(d(S_3/T)/dT)|_{T_n}$.
The PT temperature and the duration determine the peak frequency of the produced GW from PT~\cite{Huber:2008hg,Caprini:2015zlo,Caprini:2009yp}, and the trace anomaly ($\alpha$) usually determines the amplitude of the generated GW. For the $4d$ theory, the $\alpha$ is defined as
$\alpha=\Delta\rho/\rho_{rad}$
with
\begin{equation}
\Delta\rho=-\Delta V_{4d}(\phi_n,T_n)+\frac{1}{4}\left. T_n \frac{d \Delta V_{4d}(\phi_n,T)}{d T}\right|_{T=T_n}\;.\\
\end{equation}
where
$
\Delta V_{eff}(\phi,T)=V_{eff}(\phi,T)-V_{eff}(0,T)
$ and $\rho_{rad}=\pi^2g_* T_n^4/30$, $g_*=106.75$ is the effective number of relativistic degrees of freedom~\cite{Croon:2020cgk}.

To keep the perturbation of the 4d effective potential, the gauge parameter needs to satisfy the condition $|\xi|<1/g$. To ensure the $W^\pm$ mass to be positive, the chemical potential needs to be smaller than $m_W$. For these reasons, we set the $\xi=0,1$ and $\mu_{ch}=0,0.1T$ in this work. Since the difference between 2-loop order dimensional reduction and its gauge-invariant(DRGI) approach results is slight~\cite{Qin:2024dfp}, we include the DRGI results in this work to compare the differences between the 4d and 3d methods.

\begin{figure}[!htp]
    \centering
    \includegraphics[width=0.8\linewidth]{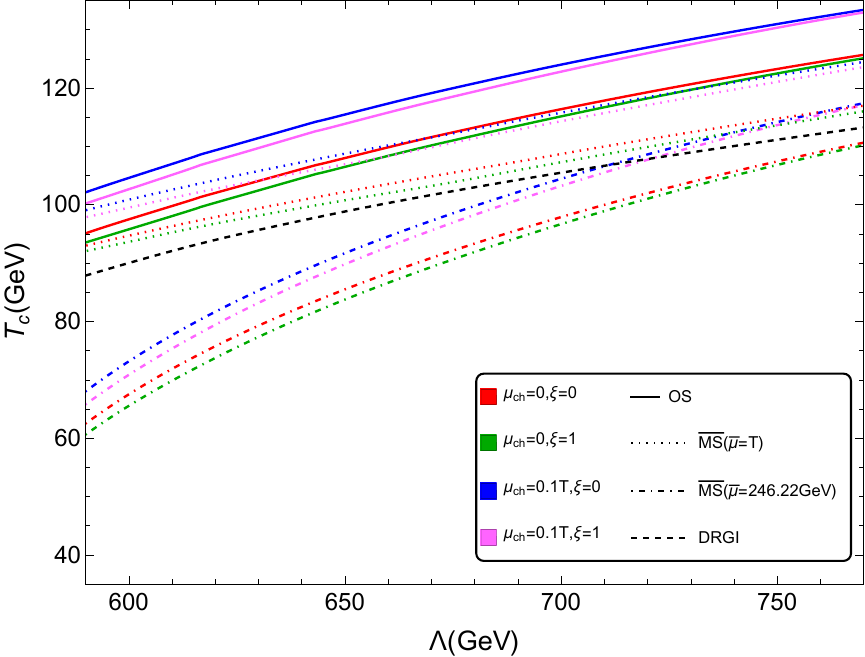}
    \includegraphics[width=0.8\linewidth]{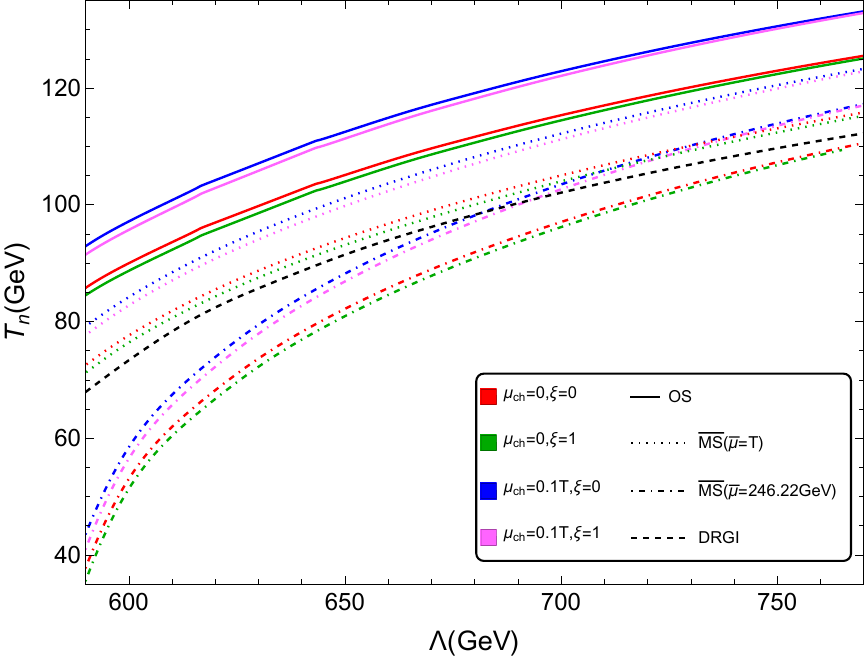}
    \caption{The temperature $T_c$(Top) and $T_n$(Bottom) as function of $\Lambda$ with the ${\overline{\rm MS}}$ and ${\rm OS}$ scheme. The solid line denotes the result of ${\rm OS}$ scheme, the dotted line denotes the result of ${\overline{\rm MS}}$ scheme with $\overline{\mu}=T$, the dot-dash line denotes the result of ${\overline{\rm MS}}$ scheme with $\overline{\mu}=246.22$GeV and the black dash line denote the result of DRGI approach. The color denotes the different values of gauge parameter and chemical potential.}
    \label{figtn}
\end{figure}

Figure~\ref{figtn} presents the dependence of both the critical temperature $T_C$ and the nucleation temperature $T_n$ on the NP scale in the ${\rm OS}$ and ${\overline{\rm MS}}$ renormalization schemes.
Notably, both temperatures follow comparable $\Lambda$-dependence patterns across the two schemes. The nucleation temperature
$T_n$ exhibits the following dependencies: 1) increases monotonically with the NP scale; 2) decreases with larger values of the gauge parameter, and 3) rises with increasing chemical potential.
Furthermore, at the fixed renormalization scale $\overline{\mu}\approx 246~\text{GeV}$,
$T_n$ is systematically higher in the ${\rm OS}$ scheme compared to the ${\overline{\rm MS}}$ scheme. This scheme dependence gradually weakens as $\Lambda$ increases.
Our analysis reveals a notable scheme dependence of the nucleation temperature $T_n$.
In the small-$\Lambda$ regime,
DRGI predicts higher $T_n$ than the
${\overline{\rm MS}}$ scheme at $\overline{\mu}\approx246~\text{GeV}$. This trend reverses in the large-$\Lambda$ regime, where the
${\overline{\rm MS}}$ yields higher $T_n$. Across the entire parameter space studied, DRGI consistently gives lower $T_n$ values compared to the ${\rm OS}$ scheme.
Furthermore, we find that the
${\overline{\rm MS}}$ scheme results at the thermal scale ($\bar{\mu}=T$) show better agreement with DRGI predictions than at other scales. Notably, the chemical potential has a stronger impact on
$T_n$ than the gauge parameter.

\begin{figure}[!htp]
    \centering
    \includegraphics[width=0.82\linewidth]{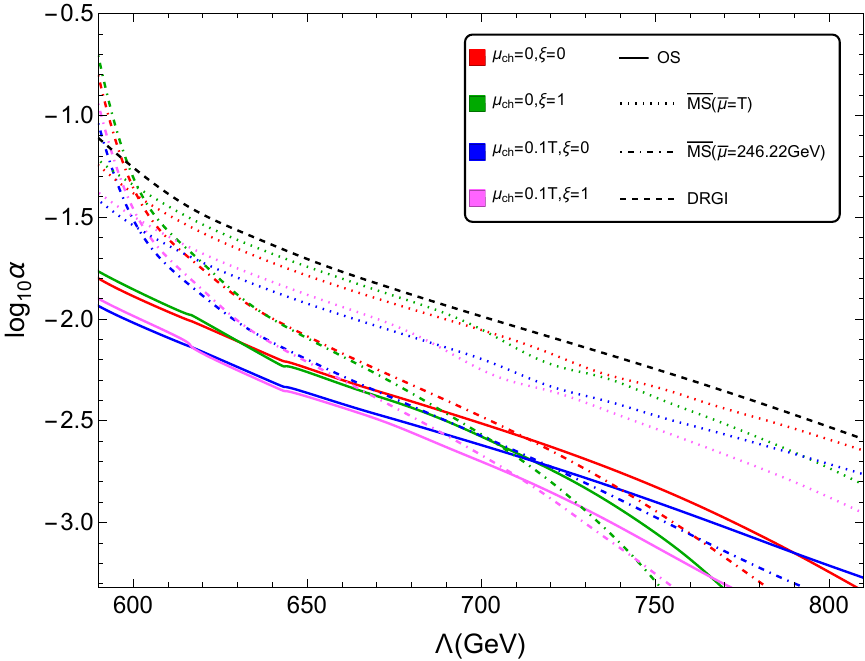}
    \includegraphics[width=0.8\linewidth]{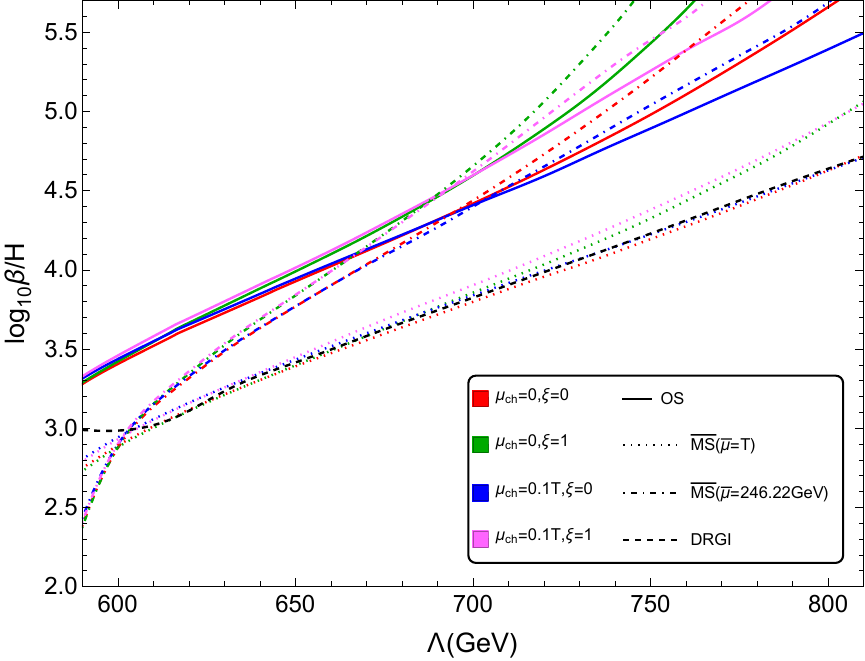}
    \caption{The parameter $\alpha$ (top) and $\beta/H$ (bottom) as functions of $\Lambda$ with the ${\overline{\rm MS}}$ and ${\rm OS}$ scheme. The color code is the same as the Fig.~\ref{figtn}.}
    \label{figalphabetah}
\end{figure}

Fig. \ref{figalphabetah} displays the dependence of $\alpha$ and $\beta/H$ on $\Lambda$ in both the ${\rm OS}$ and ${\overline{\rm MS}}$ schemes.
The parameter $\alpha$ decreases monotonically with increasing $\Lambda$.
In contrast, the $\beta/H$ exhibits a rising trend as $\Lambda$ grows. For small
$\Lambda$, an increase in the gauge parameter leads to an enhancement of $\alpha$. For large
$\Lambda$,
the same variation results in a suppression of $\alpha$. The PT strength $\alpha$ strengthens with decreasing chemical potential.
However, this effect diminishes as the NP scale $\Lambda$ increases. In the small-
$\Lambda$ regime, $\alpha$ is larger in the
${\overline{\rm MS}}$
  scheme than in the ${\rm OS}$ scheme.
Conversely, in the large-$\Lambda$ regime, this trend reverses, with the ${\rm OS}$ scheme yielding higher values of
$\alpha$. With increasing $\Lambda$, the discrepancy between the two quantities gradually decreases. For the parameter $\beta/H$, we observe that
$\beta/H$ increases with both the gauge parameter and decreasing chemical potential,
and this enhancement becomes more pronounced at larger values of $\Lambda$. In the small-$\Lambda$ regime, $\beta/H$ in ${\overline{\rm MS}}$ scheme are systematically lower than those in the ${\rm OS}$ scheme, and this trend reverses in the large-$\Lambda$ regime, where
${\overline{\rm MS}}$ yields higher $\beta/H$ values. Our analysis reveals that two-loop DRGI corrections significantly impact the phase transition parameters, as demonstrated in Refs.~\cite{Qin:2024dfp,Qin:2024idc}. Interestingly, for $\overline{\mu}=T$, the
${\overline{\rm MS}}$ scheme yields PT parameters that closely match the DRGI predictions. This agreement strongly suggests that the ${\overline{\rm MS}}$ scheme predictions display substantial sensitivity to the renormalization scale.

The differences between our results and those presented in Ref.~\cite{Ellis:2018mja} are quite pronounced, our findings indicate a higher \( T_n \) and a lower \( \alpha \).
 This discrepancy may arise from our use of the Daisy term to correct the cubic mass term in \( V_T \). In their study, the authors employed the Parwani method to modify the mass term of \( V_T \) by incorporating thermal corrections. Additionally, Ref.~\cite{Croon:2020cgk} adopts a methodology similar to ours to investigate the theoretical uncertainties associated with the phase transition using the ${\overline{\rm MS}}$ scheme. However, this paper does not include the ghost term in the one-loop effective potential, resulting in a tree-level effective potential that is not gauge invariant without the Daisy term (see Sec.\ref{secpotential}). Given that the percolation temperature $T_p$ studies there is slightly smaller than the
$T_n$ under consideration and that the difference between these two temperatures diminishes as $\Lambda$ increases, we contend that our results using the ${\overline{\rm MS}}$ scheme ($\overline{\mu}=T$) are consistent with those of Ref.~\cite{Croon:2020cgk}.

\subsection{GWs predictions}

\begin{figure}[!htp]
    \centering
    \includegraphics[width=0.8\linewidth]{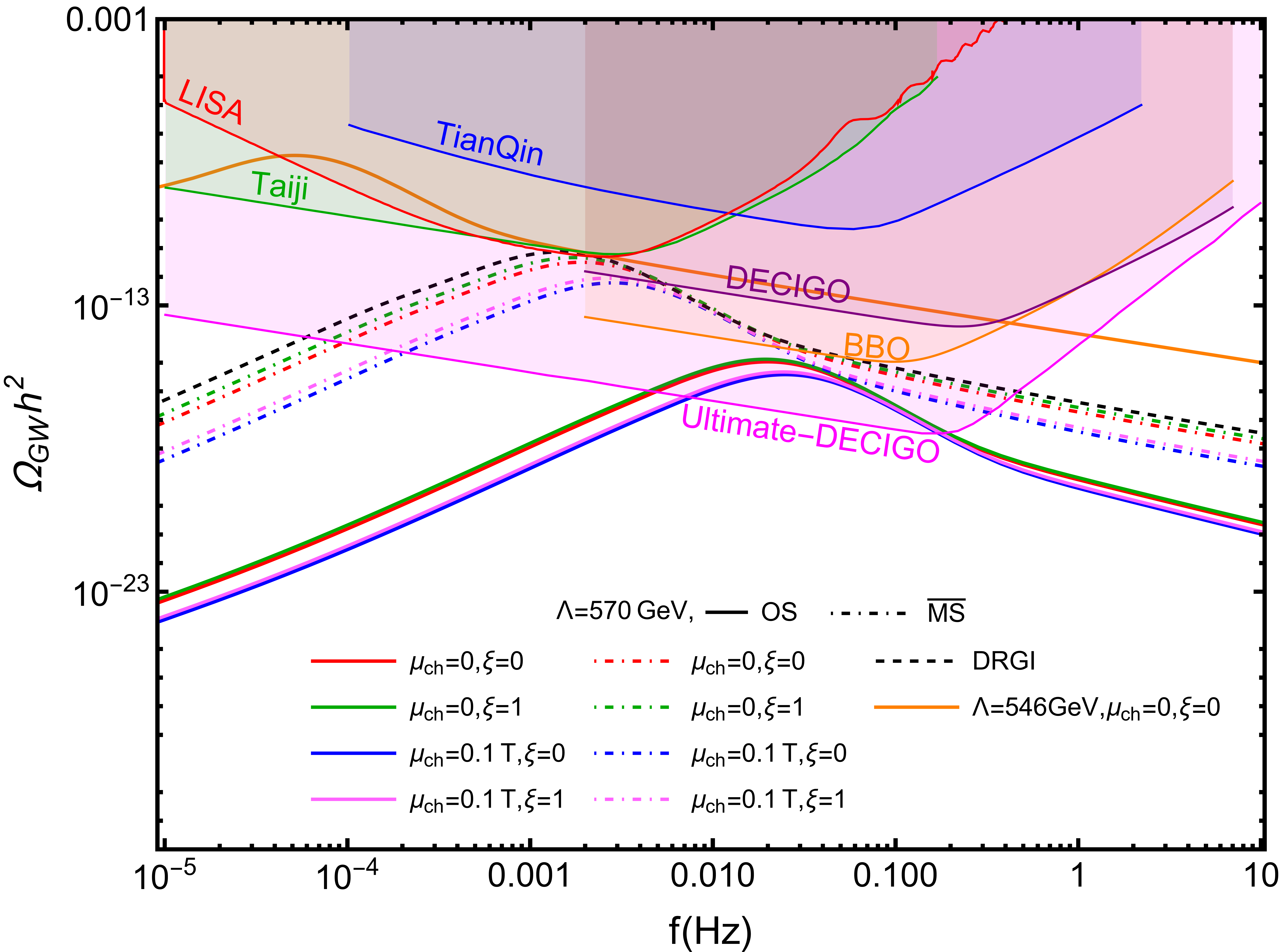}
    \caption{The effect of $\xi$ and $\mu_{ch}$ on GW predictions at $\Lambda=570$ GeV with $\overline{MS}(\overline{\mu}=T)$ and ${\rm OS}$ scheme, and the color region denotes the sensitivity of the detector. The line styles of different scenarios are the same as Fig.~\ref{figtn}.
}
    \label{figgw}
\end{figure}

For comparison, we present the predictions of GWs in the ${\rm OS}$ and $\overline{\rm MS}(\overline{\mu}=T)$ schemes in Fig.\ref{figgw}.
The figure illustrates that the strength of GWs increases with the gauge parameter while decreasing with the chemical potential. Notably, the GWs calculated within the
${\overline{\rm MS}}$ scheme at $\bar{\mu}=T$  are stronger than those obtained from the ${\rm OS}$ scheme, yet they remain undetectable by LISA for $\Lambda=570$ GeV.
Furthermore, the GWs computed using the DRGI method at the two-loop level, as proposed in Ref.\cite{Qin:2024dfp} with $\Lambda=570$ GeV, are slightly more pronounced than those predicted by the ${\overline{\rm MS}}$ scheme at one-loop when the renormalization scale is set to $\bar{\mu}=T$. Additionally, we demonstrate that GWs generated by the ${\rm OS}$ scheme at $\Lambda=546$ GeV can be detected by both LISA and Taiji, aligning with the findings of Ref.~\cite{Ellis:2018mja}.

\subsection{ Baryon Number Washout Avoidance Condition}\label{secsph}
To obtain the net baryon number, the electroweak sphaleron process needs to be suppressed at low temperature. For this reason, the parameter $\Lambda$ needs to satisfy the strongly first-order PT condition given by the baryon number washout avoidance criteria~\cite{Zhou:2019uzq,Gan:2017mcv,Qin:2024idc}):
\begin{equation}\label{PTsph}
\begin{aligned}
PT_{sph}=&\frac{E_{sph}}{T}-7\ln{\frac{v}{T}}+\ln\frac{T}{100\text{GeV}}\\
&>(35.9-42.8).
\end{aligned}
\end{equation}

Where, $E_{sph}$ is the electroweak sphaleron energy reads
\cite{Comelli:1999gt,Klinkhamer:1984di,Kuzmin:1985mm,Shaposhnikov:1987tw,Morrissey:2012db,Klinkhamer:1990fi,DeSimone:2011ek}:
\begin{equation}\label{spe3}
\begin{aligned}
&E_{sph}^{EW}(\nu,T)\\
=&\frac{4\pi v}{g}\frac{v(T)}{v}\int_{0}^{\infty}d\xi_{sph} \bigg\{\sin^2{\nu}\left(\frac{8}{3}f^{\prime 2}+\frac{4}{3}f^{\prime 2}_3\right)\\
&+\frac{4}{3}\left(\frac{g}{g^\prime}\right)^2\left[\sin^2{\nu}f_0^{\prime 2}+\sin^4{\nu}\frac{2}{\xi_{sph}^2}(1-f_0)^2\right]\\
&+\sin^4{\nu}\frac{8}{\xi_{sph}^2}\bigg[\frac{2}{3}f_3^2(1-f)^2+\frac{1}{3}(f(1-f)\\
&+f-f_3)^2\bigg]+\frac{1}{2}\xi_{sph}^2h^{\prime 2}+\frac{\xi_{sph}^2}{g^2 v(T)^4}V(h,T)\\
&+\sin^2{\nu}h^2\left[\frac{1}{3}(f_0-f_3)^2+\frac{2}{3}(1-f)^2\right]\bigg\}\;,\\
\end{aligned}
\end{equation}
when $U(1)_Y$ contribution is included.
Where, $\xi_{sph}=gvr$ and $\nu$ is related to Chern-Simons number through $N_{CS}=\frac{2\nu-\sin{2\nu}}{2\pi}$ and we set $\nu=\pi/2$. The function $f(\xi),h(\xi),f_0(\xi)$ and $f_3(\xi)$ in eq.\eqref{spe3} are the configuration of gauge and Higgs fields~\cite{Comelli:1999gt,DeSimone:2011ek,Manton:1983nd,Klinkhamer:1990fi}:
\begin{equation}\label{ansatzes2}
\begin{aligned}
g^\prime B_i d x^i&=(1-f_0(\xi))F_3 ,\\
g A_i^a\tau^a d x^i&=(1-f(\xi))(F_1\tau^1+F_2\tau^2)+(1-f_3(\xi))F_3\tau^3 ,\\
\phi&=\frac{v(T)}{\sqrt{2}}\begin{pmatrix}
                             0 \\
                             h(\xi)
                           \end{pmatrix}\;,\\
\end{aligned}
\end{equation}
with
\begin{equation}
\begin{aligned}
F_1&=-2\sin \phi d\theta -\sin 2\theta \cos \phi d\phi ,\\
F_2&=-2\cos \phi d\theta +\sin 2\theta \sin \phi d\phi ,\\
F_3&=2 \sin^2{\theta}d\phi .\\
\end{aligned}
\end{equation}
These functions  can be obtained by solving the following equations
\begin{equation}\label{dipolefuns}
\begin{aligned}
&f^{\prime \prime}+\frac{1-f}{4\xi^2}\left[8\sin^2 \nu(f(f-2)+f_3+f_3^2)\right.\\
&\left.+\xi^2h^2\right]=0,\\
&f^{\prime \prime}_3-\frac{2}{\xi^2}\sin^2 \nu\left[3f_3+f(f-2)(1+2f_3)\right]\\
&-\frac{h^2}{4}(f_3-f_0)=0 ,\\
&f^{\prime \prime}_0+\sin^2 \nu\frac{g^{\prime 2}}{4g^2} h^2(f_3-f_0)+2\frac{1-f_0}{\xi^2}=0 ,\\
&h^{\prime \prime}+\frac{2}{\xi}h^\prime-\sin^2 \nu\frac{2}{3\xi^2}h\left[2(f-1)^2\right.\\
&\left.+(f_3-f_0)^2\right]-\frac{1}{g^2v(T)^4}\frac{\partial V(h,T)}{\partial h}=0
\end{aligned}
\end{equation}
with boundary conditions:
\begin{equation}
\begin{aligned}
f(\xi)=0, h(\xi)=0, f_3(\xi)=0, f_0(\xi)=1&,\xi\rightarrow 0\;,\\
f(\xi)=1, h(\xi)=1, f_3(\xi)=1, f_0(\xi)=1&,\xi\rightarrow \infty\;.\\
\end{aligned}
\end{equation}

\begin{figure}[!htp]
    \centering
    \includegraphics[width=0.8\linewidth]{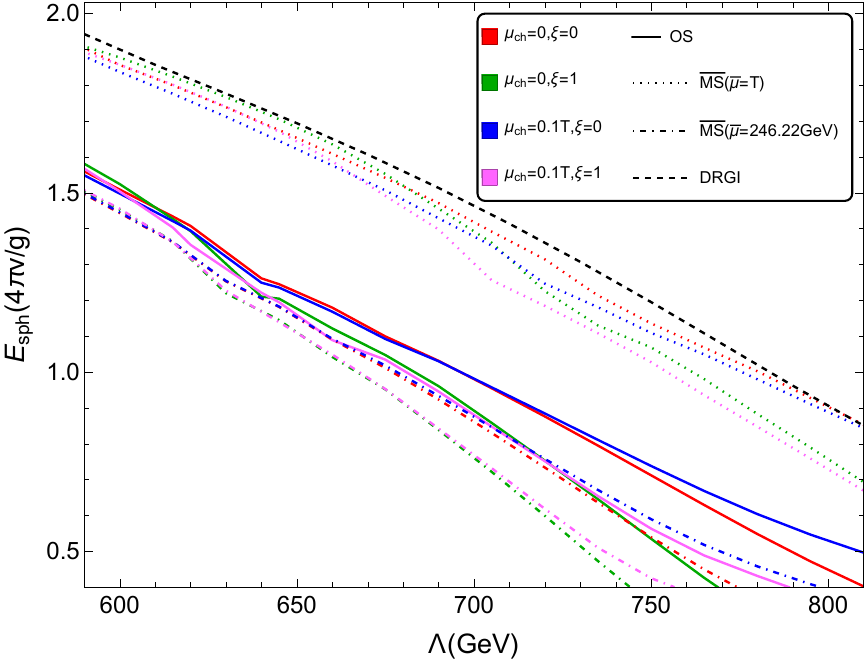}
    \includegraphics[width=0.81\linewidth]{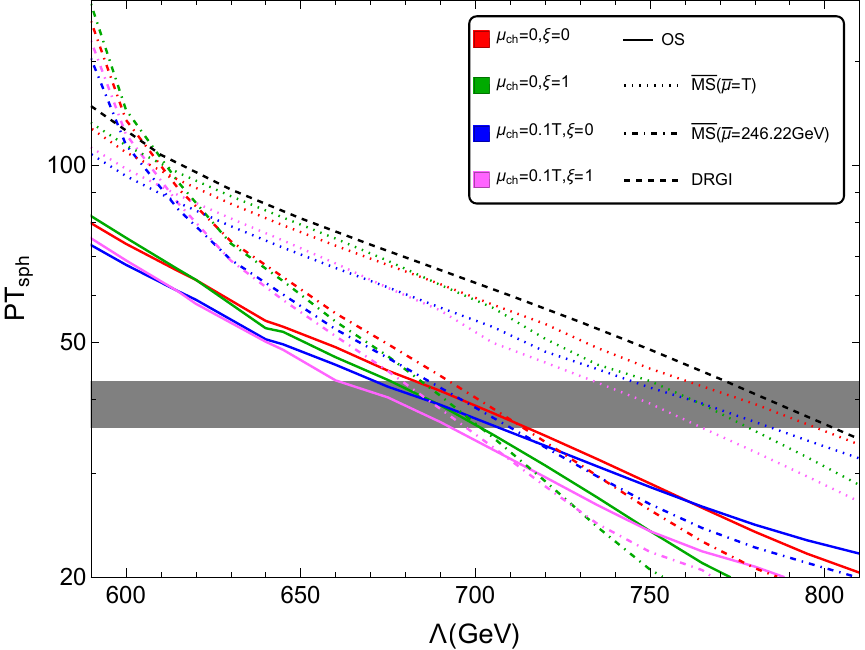}
    \caption{$E_{sph}$(Top) and $PT_{sph}$(Bottom) as function of $\Lambda$ with the ${\overline{\rm MS}}$ and ${\rm OS}$ scheme. The color codes and line styles of different scenarios are the same as Fig.~\ref{figtn}.}
    \label{figsph}
\end{figure}

Fig.\ref{figsph} quantifies the $\Lambda$ dependence of the sphaleron energy $E_{sph}$ and the value of $PT_{sph}$ under both ${\rm OS}$ and ${\overline{\rm MS}}$ schemes.
Both $E_{sph}$ and $PT_{sph}$ decrease as $\Lambda$ increases.
The $E_{sph}$ decreases with an increasing gauge parameter and increases with a rising chemical potential.
  This effect is relatively insensitive in the small $\Lambda$ region but becomes more pronounced as $\Lambda$ increases. In the
   ${\overline{\rm MS}}$ scheme(with $\overline{\mu}\approx 246~\text{GeV}$)
    $E_{sph}$ is lower than in the ${\rm OS}$ scheme; however, their behaviors are similar, and both are significantly smaller than those obtained in the DRGI scheme.
 Similarly, $PT_{sph}$ decreases with increasing gauge parameter and chemical potential. The influence of the chemical potential on $PT_{sph}$ diminishes as $\Lambda$ increases,  whereas the opposite is true for the gauge parameter. In the ${\overline{\rm MS}}$ scheme (with $\overline{\mu}\approx246~\text{GeV}$), $PT_{sph}$ is larger than in the ${\rm OS}$ scheme in the small $\Lambda$ region,
 but it becomes smaller in the large  $\Lambda$ region. The $E_{sph}$ and $PT_{sph}$ in ${\overline{\rm MS}}$ scheme (with $\overline{\mu}=T$) are lower than the DRGI results, although they exhibit similar behavior. The baryon number washout avoidance condition imposes restrictions on the model parameters. These restrictions become tighter when taking $\xi=1,\mu_{ch}=0$ for $\Lambda\lesssim680-720$ GeV in the ${\rm OS}$ scheme, $\Lambda\lesssim690-720$ GeV in ${\overline{\rm MS}}$ scheme with $\overline{\mu}\approx246$ GeV, and $\Lambda\lesssim760-800$ GeV with $\overline{\mu}=T$. These restrictions are more stringent than those provided in the DRGI approach~\cite{Qin:2024dfp,Qin:2024idc}. Additionally, the chemical potential further tightens these restrictions, while the gauge parameter exhibits a similar effect.

%The Fig.\ref{figsph} shows the effect of gauge parameter and chemical potential on the sphaleron energy at $T_n$ with different $\Lambda$. The gauge parameter has weak effect of $E_{sph}$, and it not change the limitation of $\Lambda$. The chemical potential will reduce $E_{sph}$ and $PT_{sph}$. This effect of $E_{sph}$ is more strong at large $\Lambda$ region, but it of $PT_{sph}$ is more weak at large $\Lambda$ region. This caused by the $\alpha,V_{eff}$...
%\begin{figure}[h]
%    \centering
%    \includegraphics[width=0.48\linewidth]{figsph.pdf}
%    \includegraphics[width=0.48\linewidth]{figpt.pdf}
%    \includegraphics[width=0.48\linewidth]{figsphonshell.pdf}
%    \includegraphics[width=0.48\linewidth]{figptonshell.pdf}
%    \caption{$E_{sph}$ and $PT_{sph}$ as function of $\Lambda$ with the ${\overline{\rm MS}}$(Top) and on shell(Bottom) potential.}
%    \label{figsph}
%\end{figure}

\section{Conclusion and discussion}\label{seccon}
This study presents a systematic investigation of theoretical uncertainties in perturbative calculations of EWPTs within the SMEFT framework. We comprehensively account for four key uncertainty sources: gauge dependence artifacts, chemical potential contributions, renormalization scheme selection ambiguities, and renormalization scale variations. Our findings reveal that these combined uncertainties introduce progressively larger deviations in PT parameters as the NP energy scale increases, demonstrating a positive correlation between the magnitude of BSM effects and the theoretical uncertainty budget.
Our comparative analysis reveals significant disparities between the ${\rm OS}$ and
${\overline{\rm MS}}$ renormalization schemes, particularly pronounced in SFOPT regimes with large $\alpha$.
The ${\overline{\rm MS}}$ framework exhibits pronounced scale sensitivity, manifesting as substantial theoretical uncertainties in its predictions. Notably, when implementing the baryon number preservation criterion at the EWPT temperature ($\mu=T$), the
${\overline{\rm MS}}$ scheme imposes significantly less stringent constraints on NP energy scales compared to its OS counterpart. The
${\overline{\rm MS}}$ scheme one-loop predictions for PT parameters demonstrate remarkable consistency with the results from the two-loop DRGI approach. This agreement suggests that the ${\overline{\rm MS}}$ scheme at the renormalization scale $\mu=T$ effectively encapsulates next-to-leading order corrections during thermal resummation implementation. The chemical potential (finite density effect) plays a significant role in determining the nucleation temperature. In contrast, gauge dependence introduces only modest theoretical uncertainties in perturbative calculations of phase transition PT strength and duration.

Finally, we emphasize that the validity of truncated EFT approaches in modeling first-order electroweak PTs has been critically examined in recent studies~\cite{Damgaard:2015con,Hashino:2022ghd,Postma:2020toi}. In particular, the inclusion of dimension-eight operators may become necessary for consistent calculations of the PT dynamics \cite{Hashino:2022ghd,Postma:2020toi,Chala:2018ari}. This requirement contrasts with the working regime identified in Ref.~\cite{Camargo-Molina:2024sde}, where the EFT framework remains reliable specifically when the first-order electroweak PT potential barrier originates from radiative corrections rather than tree-level contributions. In any case, we expect our observations to be broadly applicable and generalizable to other models since the theoretical uncertainties under investigation are confined to the Standard Model (SM) sector.

\section{acknowledgments}
This work is supported by the National Key Research and Development Program of China under Grant No. 2021YFC2203004, and by the National Natural Science Foundation of China (NSFC) under Grants Nos. 12322505, 12347101. We also acknowledges Chongqing Talents: Exceptional Young Talents Project No. cstc2024ycjh-bgzxm0020 and Chongqing Natural Science Foundation under Grant No. CSTB2024NSCQ-JQX0022.

%\onecolumngrid
\appendix

\section{The realtions between ${\overline{\rm MS}}$ parameters and physical observables}\label{Parameter}

We relate the $\overline{MS}-$parameters to physical observables by input Fermi constant $G_F=1.1663787\times 10^{-5}\text{GeV}^{-2}$ and pole mass\cite{ParticleDataGroup:2022pth}
\begin{equation}\label{massparameter1}
(M_t,M_W,M_Z,M_h)=(172.76,80.379,91.1876,125.1)~\text{GeV}.
\end{equation}
After use the shorthand notation $g_0^2=4\sqrt{2}G_\mu M_W^2$, the tree level relation for gauge and Yukawa couplings($g_Y$) are read
\begin{equation}\label{coulpingparameter1}
g^2=g_0^2,\quad g^{\prime 2}=g_0^2\left(\frac{M_Z^2}{M_W^2}-1\right),\quad g_Y=\sqrt{\frac{1}{2} g_0^2 \frac{M_t^2}{M_W^2}}.
\end{equation}
and $v^2=4M_W^2/g_0^2=(246.22\text{GeV})^2$ for the tree level VEV.
%We set the strong coupling constant $g_s^2=4\pi\alpha_s(m_z)$ with $\alpha_s(m_z)=0.1180$ and neglect its runing\cite{ParticleDataGroup:2022pth}.

The tree-level Higgs mass parameter and self-interaction can be obtained by solving the
\begin{equation}
\left.\frac{\partial^2 V_{tree}}{\partial \phi^2}\right|_{\phi=v_0}=M_h^2,\quad \left.\frac{\partial V_{tree}}{\partial\phi}\right|_{\phi=v_0}=0
\end{equation}
where $V_{tree}$ is defined in eq.\eqref{vefftree}. Then we can obtain
\begin{equation}
\mu_h^2=-\frac{1}{2}M_h^2+\frac{3}{4}c_6v_0^4,\quad  \lambda=\frac{1}{2}\frac{M_h^2}{v_0^2}-\frac{3}{2}c_6 v_0^2.
\end{equation}
Since the one-loop effective potential is reaching $\mathcal{O}(g^4)$, the tree-level($\mathcal{O}(g^2)$) relations need to be improved by their one-loop corrections. The relevant calculations have been done, see details in Refs.\cite{Qin:2024idc,Croon:2020cgk}.

\section{The one loop effective potential}\label{appendixoneloop}
%The explicit formula for the Coleman-Weinberg part and the finite temperature part are,
%\begin{align}
%V_{CW}=&-12V_{CW}(m_t)+6V_{CW}(m_W)+3V_{CW}(m_Z)\nonumber\\
%&+V_{CW}(m_{\chi_1})+V_{CW}(m_{\chi_2})+V_{CW}(m_{\chi_3})\nonumber\\
%&+V_{CW}(m_\phi)-2V_{CW}(m_{\eta_c})-V_{CW}(m_{\eta_0}),\\
%V_T=&12V_T(m_t/T)+6V_{T}(m_W/T)+3V_{T}(m_Z/T)\nonumber\\
%&+V_{T}(m_{\chi_1}/T)+V_{T}(m_{\chi_2}/T)+V_{T}(m_{\chi_3}/T)\nonumber\\
%&+V_{T}(m_\phi/T)-2V_{T}(m_{\eta_c}/T)-V_{T}(m_{\eta_0}/T).
%\end{align}
 The Coleman-Weinberg part and the finite temperature part has showed in Eq.\eqref{VCW}\eqref{VT}. The daisy terms of the bosonic field are defined by
 \begin{equation}
V_{Daisy}=-\sum_{i}\frac{T}{12\pi}(m_{i,res}^3-m_i^3),
 \end{equation}
with $i=\{\phi,\chi_1,\chi_2,\chi_3,W,Z,\gamma\}$, $n_{\{\phi,\chi_1,\chi_2,\chi_3,W,Z,\gamma\}}=\{1,1,1,1,2,1,1\}$, and $m_{i,res}$ is the resummed mass.
%\begin{equation}
%\begin{aligned}
%V_{Dasiy}=&V_{Daisy}(m_{\chi_1})+V_{Daisy}(m_{\chi_2})+V_{Daisy}(m_{\chi_3})\\
%&+V_{Daisy}(m_\phi)+2V_{Daisy}(m_W)+V_{Daisy}(m_Z)\\
%&+V_{Daisy}(m_\gamma).
%\end{aligned}
%\end{equation}
The field-depend masses are defined by
\begin{equation}\label{mass}
\begin{aligned}
&m_t^2=\frac{1}{2}y_t^2\phi^2,\quad m_W^2=\frac{1}{4}g^2\phi^2-\left(\mu_1-\frac{2\cos^2\theta}{\cos2\theta}\mu_3 \right)^2,\\
&m_Z^2=\frac{1}{4}(g^2+g^{\prime 2})\phi^2,\\
&m_{\eta_c}^2=\xi m_W^2,\quad m_{\eta_0}^2=\xi m_Z^2,\quad m_\gamma=0,\\
&m_{\chi_{1,2}}^2=-\mu_h^2+\lambda \phi^2+\frac{3}{4}c_6 \phi^6 +\xi m_W^2-(\mu_1-\mu_3)^2,\\
&m_{\chi_{3}}^2=-\mu_h^2+\lambda \phi^2+\frac{3}{4}c_6 \phi^6 +\xi m_Z^2-\left(\frac{\mu_3}{1-2\sin^2\theta}\right)^2,\\
&m_{\phi}^2=-\mu_h^2+3\lambda \phi^2+\frac{15}{4}c_6 \phi^6-\left(\frac{\mu_3}{1-2\sin^2\theta}\right)^2,\\
\end{aligned}
\end{equation}
With the debye mass of $A_0$ and $B_0$ field $m_D^2=\frac{11}{6} g^2T^2,m_D^{\prime 2}=\frac{11}{6} g^{\prime 2}T^2$,The resummed mass has the form
\begin{equation}
\begin{aligned}
m_{W,res}^2&=m_W^2+\Pi_W,\quad \Pi_W=m_D^2,\\
m_{Z,res}^2&=m_Z^2+\Pi_Z,\quad \Pi_Z=\frac{g^2 m_D^2+g^{\prime 2}m_D^{\prime 2}}{g^2+g^{\prime 2}},\\
m_{\gamma,res}^2&=m_\gamma^2+\Pi_\gamma,\quad \Pi_\gamma=\frac{g^{\prime 2} m_D^2+g^2 m_D^{\prime 2}}{g^2+g^{\prime 2}}.
\end{aligned}
\end{equation}
For the scalar mass parameter, we have
\begin{equation}
\mu_{h,res}^2=\mu_h^2+\Pi_\phi,\quad \lambda_{res}=\lambda+\Gamma_\lambda,
\end{equation}
with $\Gamma_\lambda=T^2 c_6$ and
\begin{equation}
\Pi_\phi=\frac{T^2}{12}(6\lambda+\frac{3}{4}(3g^2+g^{\prime 2})+3y_t^2)+\frac{1}{4}T^4 c_6\;.
\end{equation}
The resummed scalar mass are obtained by replacing the parameter $\mu_h^2, \lambda$ to $\mu_{h,res}^2, \lambda_{res}$\cite{Croon:2020cgk}.

%\begin{figure}[h]
%    \centering
%    \includegraphics[width=0.47\linewidth]{figVMS600.pdf}
%    \includegraphics[width=0.47\linewidth]{figVMS750.pdf}
%    \includegraphics[width=0.457\linewidth]{figVonshell600.pdf}
%    \includegraphics[width=0.467\linewidth]{figVonshell750.pdf}
%    \caption{The effective potential with $V_{CW}^{\overline{MS}}$(Top) and $V_{CW}^{onshell}$(Bottom) at $\Lambda=600$GeV and $750$GeV. The dash line denote the effective potential at $T_c$, and the dotdash line denote the effective potential at $T_n$.}
%    \label{figveff}
%\end{figure}

\section{Nucleation Rate}

\begin{figure}[h]
  \centering
  \includegraphics[width=0.47\linewidth]{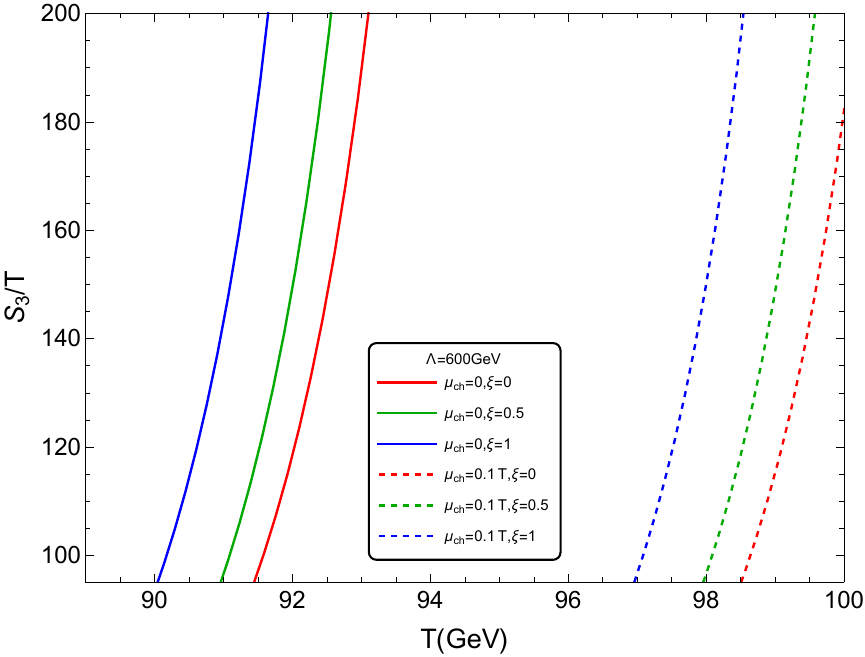}
  \includegraphics[width=0.47\linewidth]{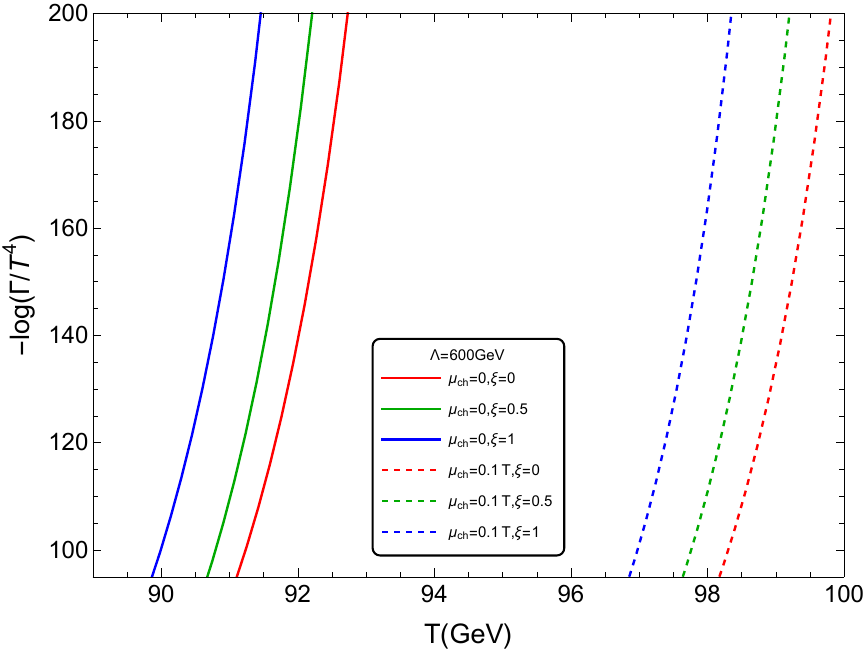}
   \includegraphics[width=0.47\linewidth]{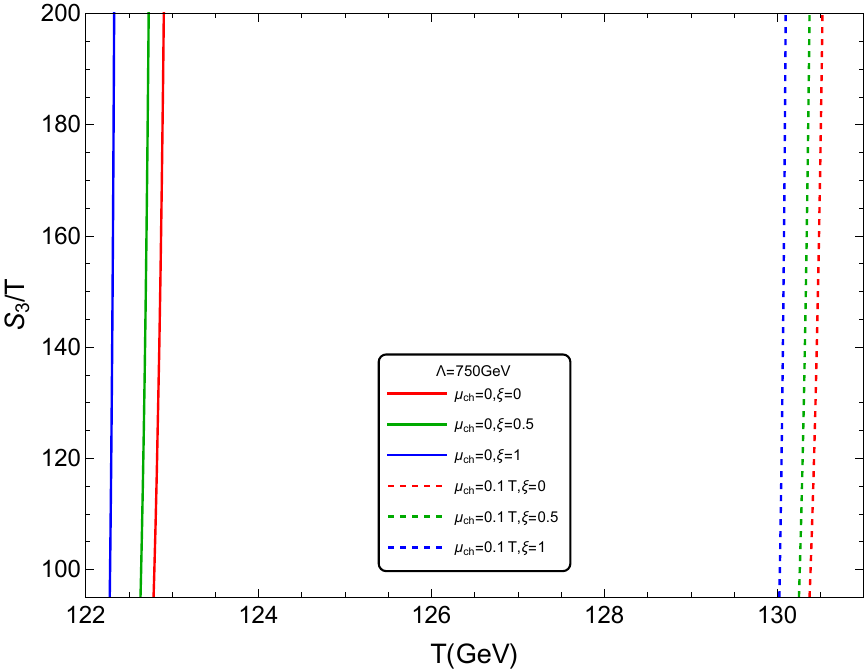}
  \includegraphics[width=0.47\linewidth]{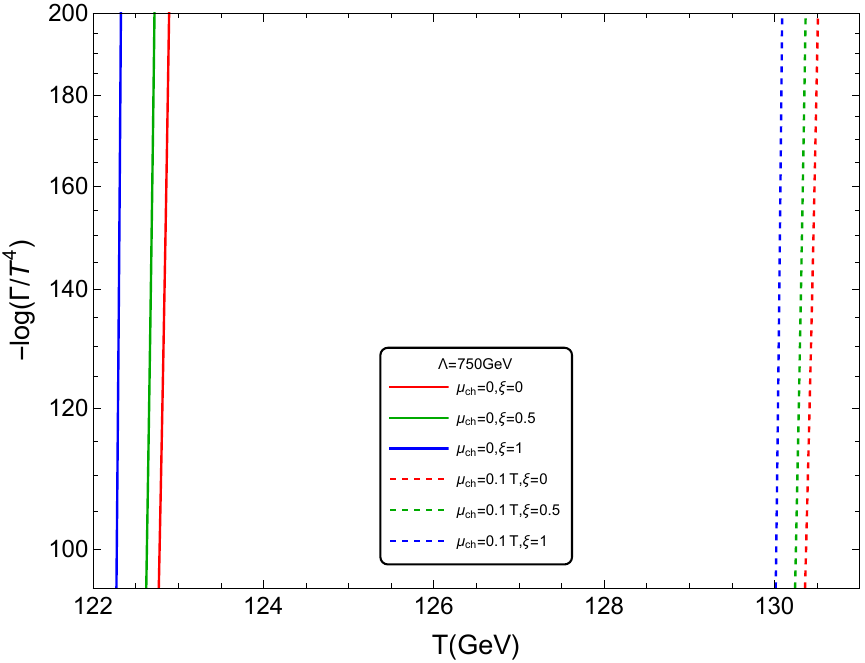}
  \caption{The $S_3/T$ and $\Gamma$ as function of temperature with $\Lambda=600$ GeV and $750$GeV at on shell potential.}
  \label{figgamma}
\end{figure}

We calculate the bubble nucleation rate by code ``bubbledet'' in this appendix\cite{Ekstedt:2023sqc}. The bubble nucleation rate is defined as
\begin{equation}
    \Gamma=A e^{-S_3/T}.
\end{equation}
The factor $A$ in finite temperature theory includes the dynamic and statistical parts
\begin{equation}
A=A_{dyn}\times A_{stat}
\end{equation}
with
\begin{equation}
\begin{aligned}
A_{dyn}&=\frac{1}{2\pi}\left(\sqrt{|\lambda_-|+\frac{1}{4}\eta^2}-\frac{\eta}{2}\right) \\
A_{stat}&=\left(\frac{S_3(\phi_b)/T}{2\pi}\right)^{3/2}\left|\frac{det^\prime(-\nabla^2+V^{\prime\prime}(\phi_b))}{det(-\nabla^2+V^{\prime\prime}(\phi_F))}\right|^{-1/2},
\end{aligned}
\end{equation}
where $\phi_F$ is the field value of false vacuum and $\phi_b$ is the bounce solution which can be obtained by solving Eq.\eqref{bouncefun}. Fig.\ref{figgamma} shows that $S_3/T$ is positively correlated with temperature, while $\Gamma$ is inversely correlated with temperature.

\section{Renormalization factor}
We calculated the Z-factor of the scalar field with $R_\xi$ gauge in this part. This factor is the coefficient of $k^2$ term in the two-point Green Function. With the definition
\begin{equation}
\begin{aligned}
&\Tint{p_b}=T\sum_{n\neq 0}\int\frac{d^d p}{(2\pi)^d},\quad \Tint{p_f}=T\sum_{n}\int\frac{d^d p}{(2\pi)^d},\\
&p_b=(\omega_b,\overrightarrow{p}),\quad p_f=(\omega_f,\overrightarrow{p}),\\
&\omega_b=2n\pi T,\quad \omega_f=(2n+1)\pi T,\\
& L_b=\ln{\frac{\overline{\mu}^2}{T^2}}-2\ln{4\pi}+2\gamma_E,\quad L_f=L_b+4\ln{2},\\
&\frac{1}{\epsilon_{b,f}}=\frac{1}{\epsilon}+L_{b,f},\quad d=3-2\epsilon,
\end{aligned}
\end{equation}
The field-dependent wave functions are
\begin{equation}
   Z=(1+Z_\phi+Z_{CT})^\frac{1}{2},\quad Z_\phi=\frac{\partial M_{hh}}{\partial k^2}.
\end{equation}
 The $M_{hh}$ can be obtained by
\begin{align}
-iM_{hh}=&2\lp\frac{i g}{2}\rp^2F_{VS}(m_W,m_{\chi_{1}})\nonumber\\
&-\lp\frac{g}{2\cos\theta}\rp^2F_{VS}(m_Z,m_{\chi_3})\nonumber\\
&+\lp\frac{2 m_W^2}{\phi}\rp^2 F_{VV}(m_W)+\frac{1}{2}\lp\frac{2 m_Z^2}{\phi}\rp^2 F_{VV}(m_Z)\nonumber\\
&+\frac{9m_h^4}{2\phi^2}F_{SS}(m_h)+\frac{m_h^4}{\phi^2}F_{SS}(m_{\chi_1})\nonumber\\
&+\frac{m_h^4}{2\phi^2}F_{SS}(m_{\chi_3})-\frac{\xi^2 m_W^4}{2\phi^2}F_{GG}(m_{\eta_c})\nonumber\\
&-\frac{\xi^2 m_Z^4}{\phi^2}F_{GG}(m_{\eta_0})-\frac{3m_t^2}{\phi^2}F_{FF}(m_t)\;,
\end{align}
where the masses are defined in Eq.\eqref{mass}. The corresponding functions are

\begin{equation}
\begin{aligned}
&F_{FF}(m)
=\Tint{p_f}\dfrac{Tr[(\slashed{p}+m)(\slashed{p}+\slashed{k}+m)]}{(p^2- m^2+i\epsilon)[(p+k)^2-m^2+i\epsilon]}\\
=&ik^2\left(-\frac{1}{8\pi^2\epsilon_f}-\frac{31m^4\zeta(5)}{256\pi^6 T^4}+\frac{35m^2\zeta(3)}{96\pi^4T^2}\right)+\cdots\;,
\end{aligned}
\end{equation}
\begin{align}
F_{VV}&(m)\nonumber\\
=&\Tint{p_b}\dfrac{\left[g_{\mu\nu}-(1-\xi)\frac{p_\mu p_\nu}{p^2-\xi m^2}\right]\left[g^{\mu\nu}-(1-\xi)\frac{(p+k)^\mu (p+k)^\nu}{(p+k)^2-\xi m^2}\right]}{\lp p^2-m^2+i\epsilon\rp[(p+k)^2-m^2+i\epsilon]}\nonumber\\
 =&ik^2\left( \frac{23 k^2 \xi ^2 \zeta (3)}{2560 \pi ^4 T^2}-\frac{49 k^2 \xi  \zeta (3)}{3840 \pi ^4 T^2}+\frac{109 k^2 \zeta (3)}{7680 \pi ^4 T^2}\right)+\cdots\;,
\end{align}
\begin{align}
F_{VS}&(m_1,m_2)\nonumber\\
=&\Tint{p_b}\dfrac{(2k^\mu+p^\mu)\left[g_{\mu\nu}-(1-\xi)\frac{p_\mu p_\nu}{p^2-\xi m_1^2}\right](2k^\nu+p^\nu)}{\lp p^2-m_1^2+i\epsilon\rp\lp(p+k)^2-m_2^2+i\epsilon\rp}\nonumber\\
=&i k^2\left(\frac{3-\xi}{16\pi^2\epsilon_b}+\frac{\xi^2 m_1^2\zeta(3)}{160\pi^4T^2}+\frac{19\xi m_2^2\zeta(3)}{1280\pi^4 T^2}\right.\nonumber\\
&\left.-\frac{47m_1^2\zeta(3)}{1920\pi^4T^2}-\frac{97m_2^2\zeta(3)}{3840\pi^4T^2}\right)+\cdots,
\end{align}
\begin{align}
F_{SS}&(m)\nonumber\\
=&\Tint{p_b}\frac{1}{(p^2-m_1^2+i\epsilon)[(p+k)^2-{m_2^2}+i\epsilon]}\nonumber\\
=&ik^2\left(\frac{ \zeta (3)}{384 \pi ^4 T^2}-\frac{ m^2 \zeta (5)}{1024 \pi ^6 T^4}\right)+\cdots,\\
F_{GG}&(m)=F_{SS}(m) \;,
\end{align}

The counterterm $Z_{CT}$ is the opposite of the coefficient of the $1/\epsilon$ term in $Z_\phi$, which serves to cancel out the divergence term $1/\epsilon$ in $Z_\phi$.

\bibliography{reference}

\end{document}